# White Dwarf Stars as Polytropic Gas Spheres

M. I. Nouh [1, 2], A. S. Saad[1, 3], W. H. Elsanhoury[1], Shaker A. A[1]., B. Korany[1,4] and T. M. Kamel[1]

[1]Department of Astronomy, National Research Institute of Astronomy and Geophysics, 11421 Helwan, Cairo, Egypt.

[2]Department of Physics, College of Science, Northern Border University, 1321 Arar, Saudi Arabia.
E-mail: abdo_nouh@hotmail.com

[3]Department of Mathematics, Preparatory Year, Qassim University, Qassim, Saudi Arabia.
E-mail: Saad6511@gmail.com

[4]Physics Dept., Faculty of Applied Science, Umm AL-Qura University, Makka, Saudi Arabia

**Abstract:** Due to the high degeneracy of electrons in compact stars, we expect that the relativistic effects play an important role in these stars. In the present article, we study the properties of the condensed matter in white dwarfs using Newtonian and relativistic polytropic fluid sphere. Two polytropic indices (namely n=3 and n=1.5) are proposed to investigate the physical characteristics of the models. We solve the Lane-Emden equations numerically. The results show that the relativistic effect is small in white dwarf stars.

## 1. Introduction

The theoretical and observational study of compact stars remains one of the most exciting fields in modern physics. Predictions of the properties of white dwarfs serve to test our understanding of matter at these high densities, while theories of high-density matter serve as a basis for interpreting observational results regarding these objects. Most exciting, these objects bring together all four of the fundamental forces of nature and probe regimes not accessible in the terrestrial laboratory (Glendening, 1996).

Matter in the interior of the compact objects is highly degenerate, and because degenerate electrons are excellent conductors of heat, the interior is nearly isothermal, and the core temperature approximately equals the temperature at the core envelope-boundary. Furthermore, because the pressure of the degenerate matter is nearly independent of the temperature, we may use the polytropic models.

Polytropic models are vital to two classes of theoretical astrophysics: stellar structure and galactic dynamics. In stellar structure, Lane-Emden equation governs the marsh of the physical variables inside configurations, Chandraseckhar (1931) and Kippenhaln and Weigert (1990).

Relativistic study of the polytropic equation of state had been done since 1964 by Tooper. Tooper (1964) derived two nonlinear differential equations analogue to the non-relativistic Lane Emden equation. By solving these two equations numerically, he could obtain the physical parameters of the polytrope. Ferrari et al. (2007) and Linares et al. (2004), solve numerically the two first order differential equations obtained by Tooper and investigated the effect of increasing a specific relativistic parameter on the polytropes with indices n=1.5 and n=3 respectively. Nouh and Saad (2013) solved TOV equation analytically at different polytropic indices.

In the present paper, we are going to study the structure of the white dwarfs using relativistic polytropic fluid spheres. The paper is organized as follows. In Section 2, polytropic and TOV equations will be discussed. Section 3 is devoted to the results reached and their interpretations. Then we pose our conclusion.

## 2. The Polytropic Gas Sphere

The polytropic equation of state has the form

$$p = K\rho^\Gamma, \qquad \Gamma = 1 + \frac{1}{n},$$

where $n$ is the polytropic index and $K$ is called the pressure constant. $\Gamma$ take 5/3 for the non-relativistic case and 4/3 for relativistic one.

The equilibrium structure of a self-gravitating object is derived from the equations of hydrostatic equilibrium. The simplest case is that of a spherical, non-rotating, static configuration, where for a given equation of state all macroscopic properties are parameterized by a single parameter, for example, the central density. By some algebraic manipulation, the structure equations could be combined to give Lane-Emden equation

$$\frac{1}{\xi^2}\frac{d}{d\xi}(\xi^2\frac{d\theta}{d\xi}) + \theta^n = 0, \qquad (1)$$

where $\theta$ and $\xi$ are dimensional and given by $\xi = rA$ and $\theta = \rho/\rho_c$. $\rho_c$ is the central density and $\rho$ is the density.

In the case of compact objects, the gravitational fields are strong enough that calculations must be performed in the context of general relativistic (rather than Newtonian) gravity. The fundamental equation of hydrostatic equilibrium in its general relativistic form has been derived by Tolman (1939) and Oppenheimer and Volkoff (1939), and is known as the "TOV" equation:

$$\xi^2 \frac{d\theta}{d\xi} \frac{1-2\sigma(n+1)\nu/\xi}{1+\sigma\theta} + \nu + \sigma \xi \theta \frac{d\nu}{d\xi} = 0, \tag{2}$$

and

$$\frac{d\nu}{d\xi} = \xi^2 \theta^n, \tag{3}$$

where

$$\nu = \frac{A^3 m(r)}{4\pi\rho_c}, \ A = \left(\frac{4\pi G \rho_c}{\sigma(n+1)c^2}\right)^{1/2}, \ \sigma = \frac{P_c}{c^2 \rho_c}, \tag{4}$$

where $\xi$ is the dimensionless radius, $\nu$ is a dimensional finite stellar mass $m(r)$ at a radius $r$, $A$ is a constant with a dimension of inverse length, $\sigma$ is the relativistic parameter (this parameter can be considered as a parameter related to the relativistic corrections) and $P_c$ is the central density. In Equations (1), (2) and (3), the Lane–Emden functions ($\theta$) are the solutions satisfying the condition $\theta = 1$ at $\xi = 0$ and $\theta = 0$ at $\xi = \xi_1$. We can determine the radius $R$ and the mass $M$ from

$$R = A^{-1}\xi_1,$$
$$M' \equiv \sigma^{(3-n)/2} \, v(\xi_1), \quad (5)$$
$$M = \frac{4\pi\rho_c}{A^3} v(\xi_1) = \left[\frac{(n+1)c^2}{4\pi G}\left(\frac{K}{c^2}\right)^n\right]^{1/2} M'.$$

## 3. Results

We integrated Equations (2) and (3) numerically using Runge-Kutta method. A Mathematica routine is elaborated to determine the zeroes of TOV equation at different polytropic indices $n$ and relativistic parameter $\sigma$. The integrations were started at initial values $\xi = 0$, $\theta = 1$, and $v = 0$ and proceeded forward using step size $\Delta\xi$. The zero of the function $\theta$, $\xi_1$, is determined by integrating until a negative value of $\theta$ is obtained. Then, the small step size $\Delta\xi$ is used to give more accurate results.

In TOV equations (Equations (2) and (3)), the functions $\theta(\xi)$, $v(\xi)$ depend on two parameters $n$ and $\sigma$. One can see that when $\sigma \to 0$ these reduce to the non-relativistic Lane-Emden equation (Equation (1)).

In Figure (1) we plot the relativistic function $v(\xi_1)$ as functions of $n$ and $\sigma$. The curve for $\sigma = 0$ reduces to the non-relativistic Lane-Emden function. The function $v(\xi_1)$ decreases with increasing $n$ as the equation of state softens and with increasing $\sigma$ as the effect of general relativity become more important.

Figure (2) plots $M'$, which determine the stellar mass, for $n \prec 3$ it is noticed $M'$ increases with increasing of $\sigma$ up to a certain maximum value which may be called the critical value $\sigma_{CR}$ and represents the onset of the instability. For $n = 3$, we can observe two minima for $M'$, namely at $\sigma = 0.4$ ($M' = 0.4516$) and $\sigma = 0.5$ ($M' = 0.4214$), while the maxima occur at $\sigma \to 0$ and $\sigma_{CR} \approx 0.42$.

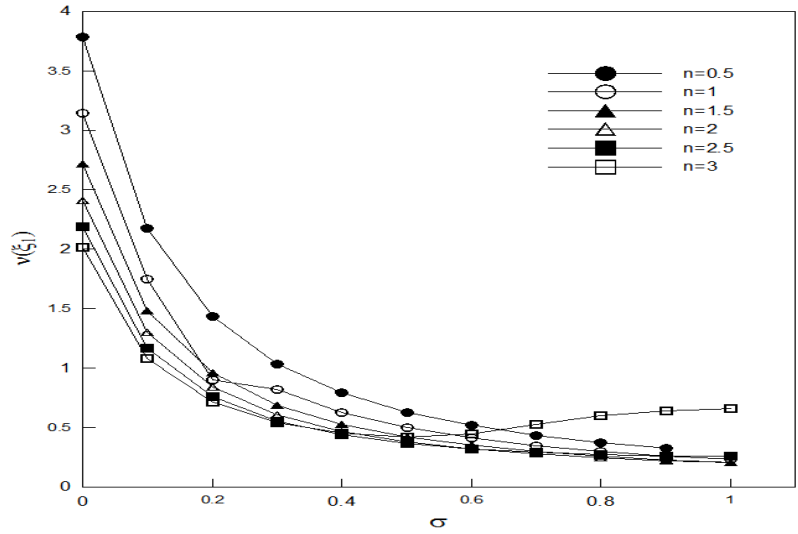

**Figure (1):** Variation of the relativistic function $\nu(\xi_1)$ with polytropic index $n$ and the relativistic parameter.

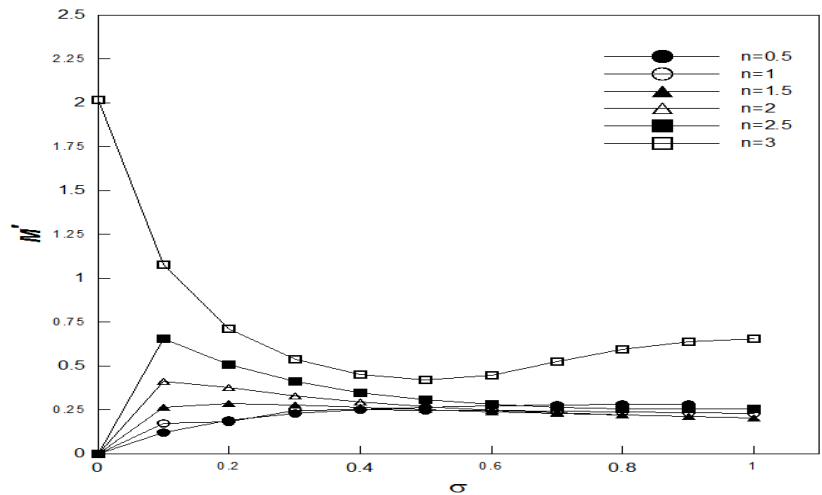

**Figure 2.** Effect of $\sigma$ on the stellar mass function $M'$.

As the matter in compact stars is highly degenerate, we expect that the relativistic effect plays very important role on the physical properties of these stars.

In Figures (3) and (4), we show the density profiles of the stellar matter for different values of $\sigma$ as a function of the radius $R(R_\circ)$. These figures show that when $\sigma$ increases, the stellar matter density is more concentrated in the center of the star. For $n=3$, the ultra-relativistic case, the effect is much stronger.

For the mass profile, the same effect as the density profiles has been obtained, since the star mass is the volume integral of the mass density. These results reflect the importance of the relativistic corrections.

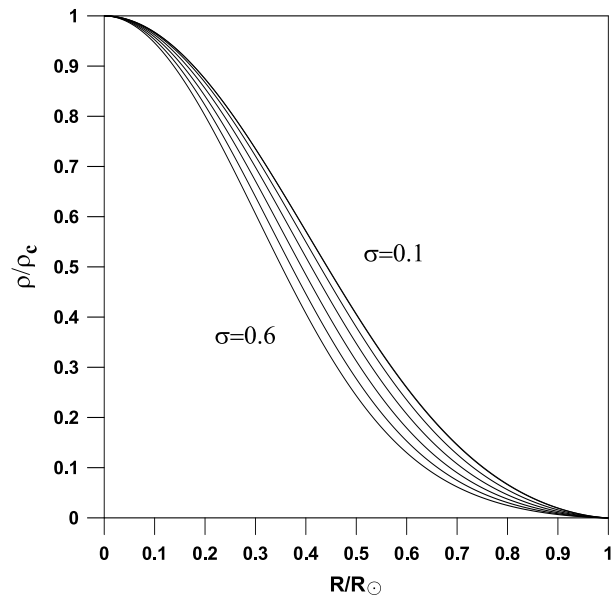

**Figure 3.** The star density profiles for different values of $\sigma$ at n=1.5.

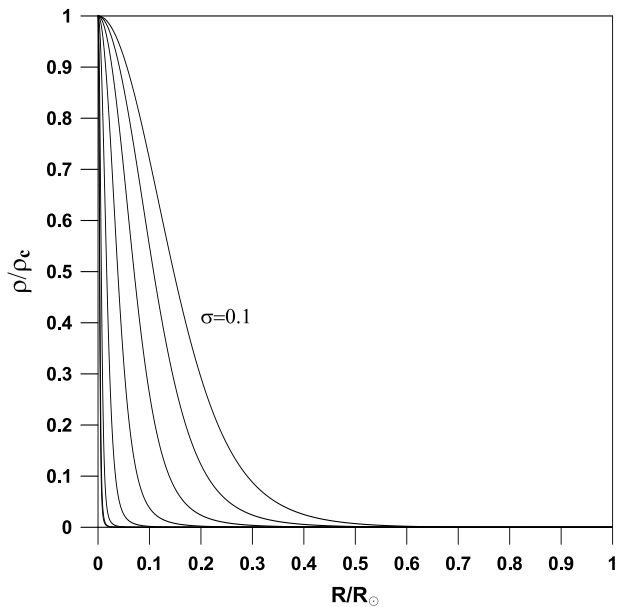

**Figure 4.** The star density profiles for different values of $\sigma$ at n=3.

Now we shall turn to the white dwarfs to determine the relativistic effect in these stars. Empirical confirmation of the theoretical mass-radius relation has been a prime objective of numerous studies employing individual stars as well as ensembles of stars with good mass and radius determinations, Holberg et al. (2012).

Nowadays, the observational projects provide masses and radii of many white dwarf stars. So, the inverse problem of white dwarfs could be established, i.e. if in the relativistic case, $M$, $R$ and $n$ are considered as a given quantities, then the determination of the relativistic parameter $\sigma$ (or range of $\sigma$) becomes a characteristic value problem and could be determined graphically, Tooper (1964).

To do this, we use the available observed mass-radius relation adopted by Provencal et al. (1998), based on the parallax of ten white dwarfs observed by HIPPARCOS. The masses and radii are listed in Table (1).

Figure (5) illustrates the position of the selected, Table (1), white dwarfs on the polytropic mass- radius relations calculated for polytropic index $n = 1.5$ and at a different value of the relativistic parameter $\sigma$. As it is seen, most of the objects tend to have small $\sigma$ except two having $\sigma$ between 0.1 and 0.3.

**Table (1):** Mass and Radii for a sample of white dwarfs (Provencal et al., 1998)

| Object's Name | M(M$_\odot$) | R(R$_\odot$) |
|---|---|---|
| Sirius B | $1.0 \pm 0.016$ | $0.0084 \pm 0.0002$ |
| Stein 2051 B | $0.48 \pm 0.045$ | $0.0111 \pm 0.0015$ |
| 40 Eri B | $0.501 \pm 0.011$ | $0.0136 \pm 0.0002$ |
| Procyon B | $0.604 \pm 0.018$ | $0.0096 \pm 0.0004$ |
| CD-38 10980 | $0.74 \pm 0.04$ | $0.01245 \pm 0.0004$ |
| W485 A | $0.59 \pm 0.04$ | $0.0150 \pm 0.001$ |
| L268-92 | $0.70 \pm 0.12$ | $0.0149 \pm 0.001$ |
| L481-60 | $0.53 \pm 0.05$ | $0.012 \pm 0.0004$ |
| G154-B5B | $0.46 \pm 0.08$ | $0.011 \pm 0.001$ |
| G181-B5B | $0.50 \pm 0.05$ | $0.011 \pm 0.001$ |
| G156-64 | $0.59 \pm 0.001$ | $0.0110 \pm 0.001$ |
| G154-B5B | $0.46 \pm 0.08$ | $0.0130 \pm 0.002$ |

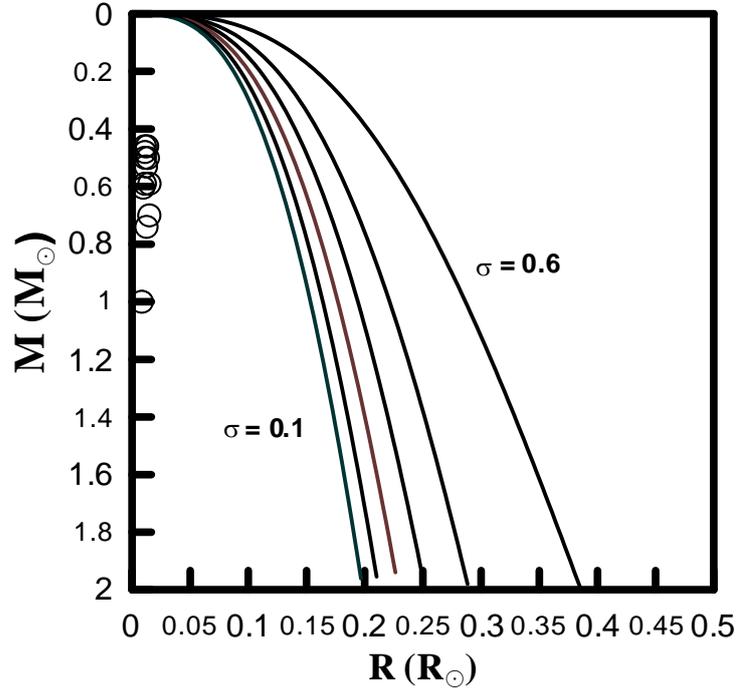

**Figure (5).** Mass radius relation for the relativistic polytrope with n=1.5. Solid lines represent mass radius relation at different relativistic parameter $\sigma$ and the open circles represent mass and radius from Provencal et al. (1998).

## 4. Conclusion

In concluding the present paper, the properties of the condensed matter in white dwarfs are studied using polytropic fluid sphere. Two polytropic indices (for n=3 and n=1.5) are considered to investigate the physical characteristics of the models. We have solved the relativistic fluid sphere equations numerically for different relativistic parameters. The deduced mass radius relation at n=1.5 is compared with observations of selected sample of white dwarfs. The result shows that the relativistic effect on the most of the selected sample is small.

## References

Binney, J. and Tremaine, S.: 1987, Galactic dynamics, Princeton Univ. Press, Princeton.